\documentclass[prl,letter,twocolumn,showpacs]{revtex4}
\usepackage{graphicx}
\usepackage{amsmath,amssymb}
\usepackage{bm}
\usepackage{ifthen}

\begin{document}
\title{Electronic phase diagrams of carriers in self-assembled InAs/GaAs
  quantum dots: violation of Hund's rule and the Aufbau principle for holes} 
\author{Lixin He}
\author{Gabriel Bester}
\author{Alex Zunger}
\affiliation{National Renewable Energy Laboratory, Golden CO 80401}
\date{\today}

\begin{abstract}

We study the orbital and spin configurations of 
up to six electrons or holes charged into
self-assembled InAs/GaAs quantum dots via
single-particle pseudopotential and many-particle configuration interaction
method. 
We find that while the charging of {\it electrons} follows both Hund's rule and
the Aufbau principle, the charging of {\it holes} follows a non-trivial 
charging pattern which violates both the Aufbau principle and Hund's rule, 
and is robust against the details of the quantum dot size. 
The predicted hole charging sequence 
offers a new interpretation of recent charging experiments.

\end{abstract}


\maketitle

The remarkable combination of three-dimensional spatial confinement in
quantum-dots with the
ability to integrate them into carrier-transporting device structures enables
storage and retrieval of electrons
\cite{drexler94,bodefeld99, regelman01,tarucha96,kouwenhoven97,
kastner93}, to the benefit of future quantum-computing 
\cite{bennett00,imamoglu99}, memory \cite{yusa97,kroutvar03}, and
single-photon \cite{michler00} applications. Unlike real atoms, 
where large
Coulomb repulsion energies $J \simeq$ 10 eV limit the number of 
ionized species to just a few, semiconductor quantum dots can be
loaded by as many as six \cite{drexler94} to 
ten \cite{banin98} electrons in colloidal \cite{banin98}
and self-assembled \cite{drexler94,bodefeld99,regelman01} dots, 
and up to hundreds of electrons in
electrostatically confined dots \cite{tarucha96,kouwenhoven97,kastner93}. 
Furthermore, one can measure for
each ionization state the stable
spin-configuration \cite{drexler94,tarucha96,kouwenhoven97,kastner93}, 
the energy to add another electron \cite{drexler94,bodefeld99,
tarucha96,kouwenhoven97,banin98} 
as well as the attendant spectroscopic shifts with charging
\cite{finley01b,regelman01}. 
Like real atoms, the stable spin configuration observed in
electrostatic dots \cite{tarucha96,kouwenhoven97,kastner93},
having lateral dimensions of 500-1000 \AA, 
follow the rules of atomic physics; that is, the 
{\it s, p, d,} ... shells are occupied in successive order with no holes 
left behind
(Aufbau principle) and with maximum spin (Hund's rule). 
Recently, it became possible to load and measure electrons
\cite{drexler94,bodefeld99} and holes
\cite{bodefeld99,reuter04,reuter05} 
into much smaller ($\simeq$ 200 x 40 \AA), 
epitaxially grown self-assembled dots of InGaAs/GaAs. 
The analysis of the results, however, was based on the effective-mass
approximation (EMA)
theoretical framework~\cite{jacak_book}, (which is
designed for dots much larger than the exciton Bohr radius), 
leading to the conclusion that electronic
configurations are atomic-like.
Electronic structure calculations for self-assembled dots
\cite{williamson00}
reveal that while for electrons the Coulomb energy 
$J_{ee} \simeq$ 20 meV is {\em smaller} than the level spacing $\Delta
\epsilon \simeq$ 50-70 meV, for holes the Coulomb repulsion  
$J_{hh} \simeq$ 15-25 meV is comparable to the level spacing 
$\Delta \epsilon \simeq$ 10-20 meV. This opens the possibility of observing
for holes some stable, exotic spin configurations 
that defy the rules of atomic physics. We
have applied a combination of an atomistic pseudopotential
description~\cite{williamson00} for the
single-particle level structure, 
with a many-body configuration interaction (CI)
description~\cite{franceschetti99}
of many-particle effects to both electron and hole loading into
InGaAs/GaAs self-assembled quantum-dots. 
We calculate the generalized electronic phase diagram of the system showing 
which many-particle configurations are energetically stable for various $p$-$p$
and $p$-$d$ splitting of the single-particle levels.
We find that while {\it electron loading}
follows both the Aufbau principle and Hund's rule, {\it hole loading}
gives rise to
stable but unusual spin configurations. 
While these calculated configurations
agree with recent measurements \cite{reuter04,reuter05}, 
they differ from their
interpretation \cite{reuter05},  which assumes 2 dimensional (2D) 
parabolic models 
\cite{warburton98} that have been
employed extensively and successfully to analyze large
electrostatically confined dots.
\cite{tarucha96,kouwenhoven97,kastner93}.
The reason for the failure of the simpler interpretation 
is that parabolic models 
ignore the inter-band and inter-valley coupling existing in a real
self-assembled quantum dot.    

\begin{figure*}
\includegraphics[width=6.2in,angle=0]{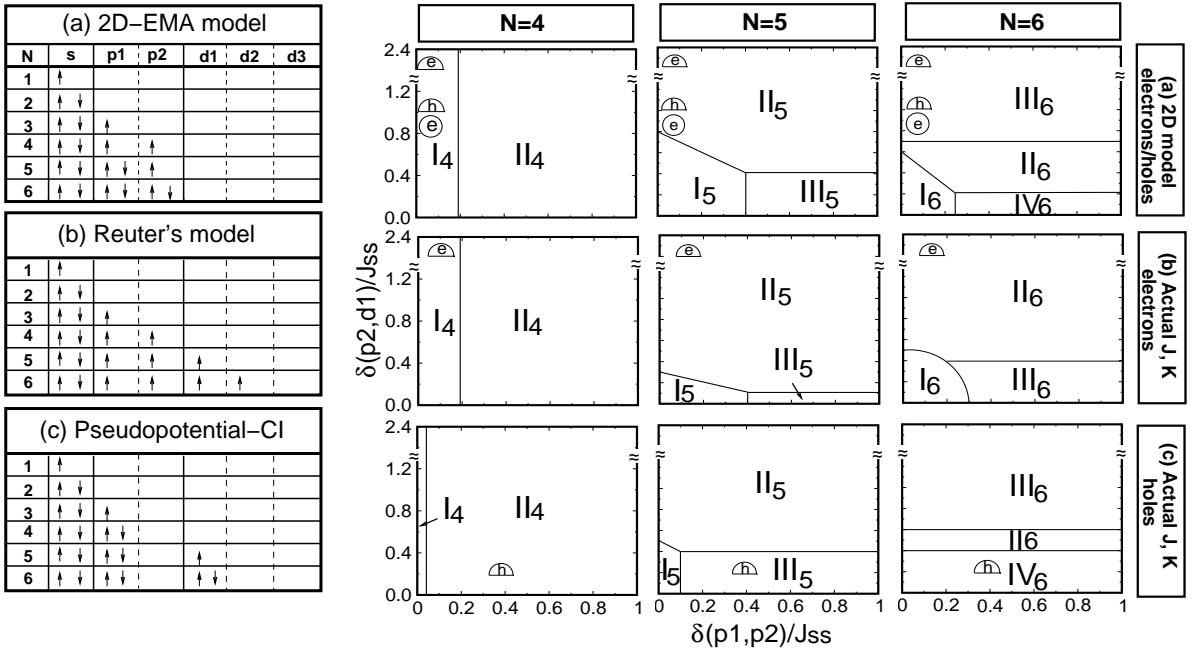}
\caption{ (Left) The ground state configurations (a) calculated from 2D parabolic
  model, (b) suggested by Reuter {\it et al} \cite{reuter05}, 
and (c) calculated from pseudopotential-CI method.
(Right)
The phase diagrams for $N=$ 4, 5, 6 electrons/holes 
(a) calculated from 2D parabolic model, 
(b) for electrons using realistic Coulomb and exchange integrals, 
and (c) for holes using realistic Coulomb and exchange integrals.
For $N$=6, $\delta_{d_1,d_2}=\delta_{p_1,p_2}$ is assumed.
The circles represent electrostatic dots, while lens represent self-assembled
dots. The labels ``e'' and ``h'' inside the symbols denote electron and hole
respectively. The configurations of phases I, II, III, and IV
 are given in the text.
}
\label{fig:combo}
\end{figure*}

The ``charging energy'' $\mu(N)$ is the energy needed to add a
carrier to the dot that is already charged by $N-1$ carriers, 
$\mu(N)=E(N)-E(N-1)$,
where $E(N)$ is the correlated many-body total energy of the ground state
of the $N$-particle dot. The ``addition energy'' $\Delta(N-1,N)$ 
(analogous to the difference
between ionization potential and electron affinity)
indicates how much more energy is needed 
to add the $N$th carrier compared to the energy needed to add the 
$N-1$th carrier:
$\Delta(N-1,N)= \mu(N)-\mu(N-1)= E(N)-2E(N-1)+E(N-2)$.
In the Hartree Fock approximation, where the effect of 
correlations is neglected but the direct Coulomb and exchange interactions
are retained, simple expressions can be derived for the addition energies.
To do so, one needs to decide first what is the order of filling the
single-particle $s$, $p$ $d$, $\cdots$ levels. 
Assuming the filling sequence of the left (L) hand side of 
Fig. \ref{fig:combo}a, 
obeying Hund's rule and the Aufbau principle, the 
addition energies are given by:
\begin{eqnarray}
\Delta_{\rm HF}(1,2)&=&J_{s,s} \; , \nonumber \\
\Delta_{\rm HF}(2,3)&=&(\epsilon_{p_1}-\epsilon_s) 
+2J_{s,p_1}- J_{s,s} -K_{s,p_1}  \; , \nonumber \\
\Delta_{\rm HF}(3,4)&=& (\epsilon_{p_2}-\epsilon_{p_1})
+ 2J_{s,p_2}-2J_{s,p_1} \nonumber \\ 
&& +J_{p_1,p_2} -K_{s,p_2} +K_{s,p_1} -K_{p_1,p_2} \nonumber \\ 
\Delta_{\rm HF}(4,5)&=& (\epsilon_{p_1}-\epsilon_{p_2}) + 2J_{s,p_1}
-2J_{s,p_2} \nonumber \\  
&&+J_{p_1,p_1}-K_{s,p_1}+K_{s,p_2}+K_{p_1,p_2}
\label{eq:add-HF}
\end{eqnarray}
where $J_{i,j}$ and $K_{i,j}$ are, respectively, 
the Coulomb and exchange integrals between
states $i$ and $j$.

To calculate these addition energies, one must first construct a
single-particle Schr\"{o}dinger equation
model.
In this step,
one might need to account not only
for quantum confinement, but also for electronic structure effects such 
as multi-band (light-hole, heavy-hole, conduction) coupling;
inter-valley ($\Gamma$-$X$-$L$) coupling; spin-orbit coupling; and the effect
of strain and chemical intermixing. 
It is then possible to compute all of 
the single-particle level spacings and integrals
entering Eq. (\ref{eq:add-HF}),
thus predict the value of $\Delta_{\rm HF} (N-1,N)$. 
Alternatively, 
one can neglect explicitly electronic structure effects other than quantum
confinement, and use instead a particle-in-a-parabolic-box model, widely used
in this field\cite{warburton98, jacak_book}. 
In this 2D-EMA, the $p$ levels are degenerate 
($\epsilon_{p_1}$=$\epsilon_{p_2}$), 
as are the $d$ levels ($\epsilon_{d_1}=\epsilon_{d_2}=\epsilon_{d_3}$)
and the splitting between the $s$ and $p$
levels ($\varepsilon_s - \varepsilon_p$) and the splitting 
($\varepsilon_p - \varepsilon_d$)
between the $p$ and $d$ levels are both
equal to the harmonic oscillator frequency $\omega$. 
Furthermore, the assumed parabolicity assures that analytic 
formulas can be derived \cite{warburton98} for the Coulomb 
and exchange matrix elements that relate all integrals
needed for the addition energies to the value of a single $J_{ss}$, 
for instance in Eq. (\ref{eq:add-HF}), $2J_{sp}-J_{ss}-K_{sp} = J_{ss}/4$. 
Thus, universal results can be derived for electrons and holes as shown in
the right (R) hand side of Fig. \ref{fig:combo}a, for $N$=4, 5, 6.
Since the restriction
of the 2D-EMA model to degenerate shells 
($\epsilon_{p_1}$=$\epsilon_{p_2}$, 
$\epsilon_{d_1}=\epsilon_{d_2}=\epsilon_{d_3}$ ) 
and to equidistant
shells ($\epsilon_p-\epsilon_s=\epsilon_d-\epsilon_p$) 
might be rather stringent\cite{footnote1}, we allow 
in Fig. \ref{fig:combo}a(R),
$\delta_{p_1,p_2}=\epsilon_{p_2}-\epsilon_{p_1}$ and 
$\delta_{p_2,d_1}=\epsilon_{d_1}-\epsilon_{p_2}$ to vary, calculating for each 
$\{ N; \delta_{p_1,p_2}, \delta_{p_2,d_1} \}$ 
the configuration which minimizes
the total-energy. This gives a phase diagram as a function of
the parameters $\delta_{p_1,p_2}$ and $\delta_{p_2,d_1}$ 
in the unit of $J_{ss}$ as shown in Fig. \ref{fig:combo}a(R) for particle
number $N$=4, 5 , 6.
The 2D-EMA model yields for $N$=4 two electronic phases: high-spin
I$_{4}$=$(s^\uparrow s^\downarrow)(p_{1}^{\uparrow})(p_{2}^{\uparrow})$
and low-spin II$_{4}$=$(s^\uparrow s^\downarrow)
(p_{1}^{\uparrow} p_{1}^{\downarrow})$.
For $N=5$, we find three phases I$_{5}$=
$(s^\uparrow s^\downarrow)(p_{1}^{\uparrow})
(p_{2}^{\uparrow})(d_1^{\uparrow})$, II$_{5}$=
$(s^\uparrow s^\downarrow)(p_{1}^{\uparrow}p_{1}^{\downarrow})
(p_2^{\uparrow})$ and III$_{5}$=
$(s^\uparrow s^\downarrow)(p_{1}^{\uparrow}p_{1}^{\downarrow})
(d_{1}^{\uparrow})$.
For $N=6$, we find four phases, I$_{6}$=
$(s^\uparrow s^\downarrow)(p_{1}^{\uparrow})
(p_{2}^{\uparrow})(d_1^{\uparrow})(d_2^{\uparrow})$, 
II$_{6}$=
$(s^\uparrow s^\downarrow)(p_{1}^{\uparrow}p_{1}^{\downarrow})
(p_2^{\uparrow})(d_1^{\uparrow}) $, III$_{6}$=
$(s^\uparrow s^\downarrow)(p_{1}^{\uparrow}p_{1}^{\downarrow})
(p_{2}^{\uparrow}p_{2}^{\downarrow})$ and IV$_{6}$=
$(s^\uparrow s^\downarrow)(p_{1}^{\uparrow}p_{1}^{\downarrow})
(d_{1}^{\uparrow}d_{1}^{\downarrow})$.
To decide which of these phases is a ground state, we need to know in 
Fig. \ref{fig:combo}a(R)
the value of $\delta_{p1,p2}/J_{ss}$ and $\delta_{p2,d1}/J_{ss}$. 
For electrons in self-assembled dots, the single-particle energy spacing
is usually more than twice the Coulomb energy, so,
$\delta_{p2,d1} > 2 J_{ss}$ \cite{drexler94,warburton98}. 
For holes, $\delta_{p2,d1}=1.17 J_{ss}$ was determined 
from recent experiments\cite{reuter04, reuter05}
and $\delta_{p1,p2}=0$  is assumed in 2D-EMA model.
This places in Fig. \ref{fig:combo}a(R) 
for both electrons and holes, phases
I$_4$, II$_5$, III$_6$ as ground states for $N$=4, 5, 6, respectively.
The ground state configurations of the 2D-EMA model are collected in 
Fig. \ref{fig:combo}a(L), for $N$=1 - 6. 

\begin{table*}
\caption{
Hole addition energies of self-assembled InAs/GaAs quantum dots
in meV. The experimental values are from
Ref. \cite{reuter05} at zero magnetic field. The ``Ground State'' values
correspond to the low-spin configurations as given in Fig.~\ref{fig:combo}a(L)
and the ``Excited State'' values to the high-spin configurations assumed in
Ref. \onlinecite{reuter05} and given in Fig.~\ref{fig:combo}b(L). The results
of ``Pseudopotential+CI'' calculations correspond to the configurations from 
Fig.~\ref{fig:combo}c(L).
}
\label{tab:additionE}
\vskip 0.2cm
\begin{tabular}
{l| c| c c| cc cc cc}
\hline\hline
Addition  & Exp.  &
\multicolumn{2}{c|}{2D-EMA model} &
\multicolumn{6}{c}{Pseudopotential + CI} \\
Energy & & Ground state & Excited state &
\multicolumn{2}{c}{2$R$=20 nm} &
\multicolumn{2}{c}{2$R$=25 nm}&
\multicolumn{2}{c}{2$R$=27.5 nm}\\ 
&& Fig.\ref{fig:combo}a(L)  & Fig.\ref{fig:combo}b(L) 
&$h$=2.5 nm & $h$=3.5 nm & $h$=2.5 nm & $h$= 3.5 nm 
& $h$=2.5 nm & $h$= 3.5 nm\\
\hline
$\Delta_h(1,2)$ &23.9 &Fitted&Fitted& 
24.1 & 19.0  & 21.9 & 17.5 & 21.0 & 16.7\\
$\Delta_h(2,3)$ &34.2 &Fitted&Fitted& 
28.7 & 21.7  & 27.2 & 21.2 & 26.4 & 20.6\\
$\Delta_h(3,4)$ &17.1 &12&12& 18.1 & 16.9  & 16.4 & 15.2 & 15.6 & 14.5\\
$\Delta_h(4,5)$ &23.2 &21&29& 26.4 & 21.6  & 25.4 & 20.8 & 23.8 & 20.5\\
$\Delta_h(5,6)$ &15.0 &12&18& 17.1 & 16.1  & 15.3 & 14.4 & 15.5 & 13.7\\
\hline\hline
\end{tabular}
\end{table*} 

For {\it electrons}, the ground states of 2D-EMA model 
are corroborated by atomistic
pseudopotential calculations [Fig.\ref{fig:combo}b(R)], 
where we
use the Coulomb integrals obtained from atomistic 
wavefunctions for electrons in
a lens shape InAs/GaAs 
dot with 25 nm base and 2.5 nm height. 
This shape is realistic,
according to experimental findings~\cite{walther01}, 
and predicts a fundamental photoluminescence line very close to
the one observed in the charging experiment \cite{reuter05} at around 1 eV.
Overall, the comparison between Figs. 2a and Figs. 2b 
shows that while the phase boundaries 
can change significantly when realistic wavefunctions are assumed instead of
2D-EMA values, the ground state symmetries 
for N=5 and 6 electrons in self-assembled dots remain unchanged and 
are far from other competing phases.

The foregoing analysis of loading of {\em electrons}
\cite{tarucha96,kouwenhoven97,kastner93} 
has been simplified by the fact
that the single-band effective mass model is not a drastic approximation
given that in direct-gap zinc-blende materials electrons derive from a
nondegenerate, spin-orbit-free $\Gamma_{1c}$ band which is energetically
isolated from other states. The analysis of loading of 
{\em holes}, however, does
not benefit from the same simplification, as holes derive from a mix of heavy-
and light-hole states,
invalidating \cite{he04a} the classification of hole states as pure 
$s$ or $p$ or $d$ levels and as pure heavy-hole or light-hole states
\cite{footnote2}. 
Furthermore, as shown in Fig. \ref{fig:combo}a(R), 
{\it unlike} electrons, the ground states
of holes are close to competing phases in the 2D-EMA phase diagrams, which 
require a more careful treatment.
Nevertheless, the 2D-EMA model is still attractive in its simplicity and 
Reuter {\em et al.} \cite{reuter05} used it to analyze their hole charging
results. 
The value of $J_{ss}$
is directly accessible from experiments since 
it is well approximated by $\Delta(1,2)$.
The only remaining parameter in the 2D-EMA model  
is the single-particle energy splitting 
$(\epsilon_s - \epsilon_p)$=$(\epsilon_p - \epsilon_d)$=$\omega$ 
which can be extracted from measuring 
$\Delta(2,3)=(\epsilon_p-\epsilon_s)+J_{ss}/4$.
Reuter {\it et al} \cite{reuter05} thus determined $J_{ss}$= 23.9 meV
and $\varepsilon_p - \varepsilon_s$ = 28 meV.
Since experimentally five addition
energies are available and only two were used in the fit above, 
the problem is overdetermined and it is possible
to assess how well the model reproduces the remaining 
three experimental data points.
Assuming $\delta_{p1,p2}$=0 and $\delta_{p2,d1}$=$\epsilon_{d} -\epsilon_p$
=$\epsilon_{p} -\epsilon_s$=28 meV, yields the above mentioned
$\delta_{p2,d1}$=1.17 $J_{ss}$,
leading to the phase diagram of Fig. \ref{fig:combo}a(R), 
with ground state phases: I$_4$,
II$_5$ and III$_6$. If one calculates 
the addition energies of 
Eq.(\ref{eq:add-HF}), using these ground-state configurations
one gets the values indicated by ``2D-EMA model ground state'' in Table 
\ref{tab:additionE}. 
However, this hole addition sequence contradicts the magnetic field data of
Ref. \onlinecite{reuter05},
which show that the hole $d$-levels are occupied
before $p$-levels are filled completely (non-Aufbau) \cite{reuter05}.  
To explain their magnetic field data, 
Reuter {\it et al.} \cite{reuter05} 
assumed a hole filling sequence 
[Fig.~\ref{fig:combo}b(L)] that relies on an {\it ad hoc} excited hole state
instead of the the ground states
predicted by the 2D-EMA model [Fig.~\ref{fig:combo}a(L)].
In Table \ref{tab:additionE}, we compare the ensuing addition energies
of both the ground state and excited state configurations 
with experiments. 
We find that the addition energies given by 
both scenarios of Fig. \ref{fig:combo}a(L) and \ref{fig:combo}b(L)
show significant discrepancies 
from the experimental values, with about 25 -50 \% error.

For holes, the discrepancy between the harmonic oscillator 
results Fig.~\ref{fig:combo}a(L)
(Aufbau configuration)
and the magnetic field data of Fig.~\ref{fig:combo}b(L)  
(non-Aufbau configuration)
is difficult to reconcile within the 2D-EMA model,  
since even relative large variations of the single-particle energy parameters
$\delta_{p1,p2}$ and 
$\delta_{p2,d1}$ do not lead to a ground-state configuration change from
Fig.\ref{fig:combo}a(L) to Fig.\ref{fig:combo}b(L). 
Our atomistic pseudopotential, plus CI
calculations show different ground-state configurations for holes than the
above two models. 
As shown in Fig.\ref{fig:combo}c(R), 
for $N$=4, and $N$=5, using an atomistic description
the topology of the phase-diagrams are the
same as in the 2D-EMA model,
but the boundaries are shifted.
As a result, for $N$=4, the
hole ground state is now phase II$_{4}$=$(s^\uparrow s^\downarrow)
(p_{1}^{\uparrow} p_{1}^{\downarrow})$, not 
I$_{4}$=$(s^\uparrow s^\downarrow)(p_{1}^{\uparrow})(p_{2}^{\uparrow})$ ,
and for $N$=5, the ground state is phase III$_{5}$=
$(s^\uparrow s^\downarrow)(p_{1}^{\uparrow}p_{1}^{\downarrow})
(d_{1}^{\uparrow})$, not II$_{5}$=
$(s^\uparrow s^\downarrow)(p_{1}^{\uparrow}p_{1}^{\downarrow})
(p_2^{\uparrow})$.
For $N$=6, the topology of the phase digram changed completely: phase
I$_{6}$=$(s^\uparrow s^\downarrow)(p_{1}^{\uparrow})
(p_{2}^{\uparrow})(d_1^{\uparrow})(d_2^{\uparrow})$ disappeared, 
and the ground state is now IV$_{6}$=
$(s^\uparrow s^\downarrow)(p_{1}^{\uparrow}p_{1}^{\downarrow})
(d_{1}^{\uparrow}d_{1}^{\downarrow})$. The ground-state
configurations are listed in Fig.\ref{fig:combo}c(L) 
for $N$=1 - 6.  
Using these new ground states, Table \ref{tab:additionE} compares
the experimental addition energies 
and the calculated results for six different 
InAs/GaAs lens shaped quantum dots of different bases and heights. 
Very good agreement is obtained for the InAs dot with 
20 nm base and 2.5 nm height, with differences in the 
addition energies of less than 16\%,
compared with almost 50\% error in 2D-EMA model (despite the fact that
two of the addition energies were fitted).
The parameters $\delta_{p1,p2}$ and $\delta_{p2,d1}$ calculated for 
different dots are given in Table \ref{tab:deltaE} and as shown in 
Fig. \ref{fig:combo}c(R), lie close to the center 
of the predicted phases II$_4$, III$_5$ and IV$_6$. 
This indicates the stability
of our numerical results against possible variations of $\delta_{p1,p2}$ and 
$\delta_{p2,d1}$ due to shape anisotropy or alloy effects. 
The addition energies we calculate from configuration interaction
are different by about 1-3 meV from those we calculate in Hartree-Fock, 
although they give the same ground state.

\begin{table}
\caption{First hole Coulomb energy $J_{ss}$ and
single-particle energy level spacings in meV, from 
atomistic pseudopotential calculations for six different self-assembled
len-shaped InAs/GaAs quantum dots.}
\label{tab:deltaE}  
\vskip 0.2cm
\begin{tabular}{l cc cc cc}
\hline\hline
&\multicolumn{2}{c}{2$R$=20 nm} &
\multicolumn{2}{c}{2$R$=25 nm}&
\multicolumn{2}{c}{2$R$=27.5 nm}\\  
                  &$h$=2.5&$h$=3.5&$h$=2.5&$h$= 3.5&$h$=2.5&$h$= 3.5\\ 
\hline
J$_{ss}$         &  27.2 & 22.1  & 25.1  & 20.4   & 24.2  & 19.6 \\
$\delta_{p1,p2}$ & 10.9  & 11.3  & 7.1   & 9.5    & 5.8   & 7.9 \\
$\delta_{p2,d1}$ &  4.5  & 3.4   & 8.34  & 2.4    & 9.4   & 3.9\\
\hline\hline
\end{tabular}
\end{table} 

Our predicted charging pattern [Fig. \ref{fig:combo}c(L)] 
shows that the level filling by holes
does not follow the Aufbau principle nor the Hund's rules:
$d$ levels get filled {\rm before} the second $p$ level, despite the fact
that the $d$ level is energetically more than 3 meV above the second
$p$ level.
The non-trivial hole filling pattern is due
to two reasons. First, the large $p$-level splitting leads to
the $p_2$ level being energetically 
close to the $d_1$ level, i.e., $\delta_{p_2,d_1}$ is small.
In Table~\ref{tab:deltaE}, we list the hole single-particle energy spacings
and the first hole Coulomb integrals $J_{ss}$ for different dots. We see
that $\delta_{p_1,p_2} \sim$ (0.3 - 0.5) $\cdot J_{ss}$ and
$\delta_{p_2,d_1} \sim$ (0.2 - 0.3)  $\cdot J_{ss}$, which differ
significantly from the assumption of 2D-EMA where, 
$\delta_{p1,p2} \sim$ 0, and $\delta_{p2,d1} \sim J_{ss}$.
Second, the Coulomb repulsion between the $p_1$ and the $d$ level  
is lower than that 
between the two $p$ levels, therefore the Coulomb energy can 
overcome the single-particle energy spacing $\delta_{p_2,d_1}$, leading to
the non-Aufbau charging pattern. 

An important feature of the present theory is not only 
its compatibility with the 
zero field experimental results but also with the magnetic field dependence
obtained in Ref. \onlinecite{reuter05}. 
The experimental results from Ref. \onlinecite{reuter05} 
show that the spin Zeeman effect is significantly 
smaller than the orbital Zeeman and in the interpretation of the experimental 
results only the latter has to be taken into account. 
Within this approximation  our theoretical results reproduce 
the experimental data for $N$ = 1 - 6 holes given in  Fig.~2 of 
Ref. \onlinecite{reuter05},
since our results have the same total orbital angular momenta.
The difference in the total spin for configurations assumed in 
Ref. \onlinecite{reuter05} and in the present work is significant. 
We suggest to identify the spin configurations by measuring optically
the fine-structure of the $N\, h +e \rightarrow (N-1)\,h$ transition 
in large magnetic field. While our suggested configurations
for the $N$ = 4, 6 holes have closed-shell 
fillings and only one single line with no  fine-structure 
should be observed, the high-spin configuration suggested by Reuter 
{\em et al.} \cite{reuter05} should show a rich fine-structure that should
become apparent, especially at high magnetic fields.

In conclusion, we analyzed the many-particle configurations for
electrons and holes in quantum dots in the form of
phase diagrams. From these diagrams, we predict that the hole charging
sequence presents surprising configurations (not expected from effective
mass calculations) that violates the Aufbau principle
as well as the Hund's rule. 
Our results are in good agreement with recent experimental findings and
provide a novel way to study the charging of carriers in quantum dots.

\acknowledgments
This work was supported by the US Department of Energy, Office of Science,
Basic Energy Science.


\begin{thebibliography}{24}
\bibitem[{foo({\natexlab{a}})}]{footnote1}
\bibinfo{note}{Indeed, there are indications that the $p$-levels and the
  $d$-levels of self assembled quantum dots are, in fact split. This is known
  from multi-excitons measurements \cite{regelman01} as well as from our
  theoretical analysis \cite{bester05a}, showing that the underlying
  zinc-blende symmetry induces such splitting $\epsilon_{p_1}\neq
  \epsilon_{p_2}$ and $\epsilon_{d_1}\neq \epsilon_{d_2}$, even for
  cylindrically symmetric dots.}
\bibitem[{foo({\natexlab{b}})}]{footnote2}
\bibinfo{note}{For example, the first 3 electron and hole wavefunctions of
  lens-shaped InAs/GaAs quantum dot with base 2$R$=25 nm and height $h$=3.5 nm
  are plotted in Ref.\onlinecite{he04a}, where the projection of wave functions
  into angular momentum are listed in Table II of Ref.\onlinecite{he04a}. As
  can be seen both electron and holes states are not pure $s$ or $p$ unlike
  what is assumed in the 2D-EMA.}
\end{thebibliography}
\end{document}